\documentclass[draft,english,aps,prb]{revtex4}
\usepackage[T1]{fontenc}
\usepackage[latin1]{inputenc}
\usepackage{amsmath}
\usepackage{amssymb}

\makeatletter

\DeclareMathOperator{\arctanh}{arctanh}

\usepackage{babel}
\makeatother
\begin{document}
\begin{abstract}
Relativistic rapidity is usually presented as a computational device.
As L{\'e}vy-Leblond has shown, it is also the velocity that would
be imputed by an ideal Newtonian inertial guidance system, taking
$c=1\,\textrm{neper=1}$.  Here, we show that it can also be interpreted
as the change in musical pitch of radiation fore and aft along the
direction of motion.
\end{abstract}

\title{Relativistic rapidity as change in musical pitch}

\author{Alma Teao Wilson}

\maketitle
Department of Mathematics\\
Brigham Young University\\
Provo, Utah, USA

\section{Introduction}

The usual one-dimensional relativistic velocity addition formula is
given by\begin{equation}
\frac{v_{\textrm{AB}}}{c}\boxplus\frac{v_{\textrm{BC}}}{c}=\frac{\frac{v_{\textrm{AB}}}{c}+\frac{v_{\textrm{BC}}}{c}}{1+\frac{v_{\textrm{AB}}}{c}.\frac{v_{\textrm{BC}}}{c}}.\label{eq:vel-addn}\end{equation}

The rapidity $\alpha$ for a velocity $v$ is defined by \begin{equation}
\tanh(\frac{\alpha}{b})=\frac{v}{c},\label{eq:tanh-rapidity -is-velocity}\end{equation}

with $b$ an arbitrary nonzero constant usually chosen to be unity.
We find that\begin{equation}
\alpha_{\textrm{AB}}+\alpha_{\textrm{BC}}=\alpha_{\textrm{AC}}\Leftrightarrow\frac{v_{\textrm{AB}}}{c}\boxplus\frac{v_{\textrm{BC}}}{c}=\frac{v_{\textrm{AC}}}{c}.\label{eq:rapidity-addition}\end{equation}

In other words, relativistic velocity addition in one spatial dimension
is equivalent to ordinary addition of the corresponding rapidities.
In a companion paper\cite{Wia07velmult}, we discuss the use of velocity
factors (two-way doppler factors) to simplify calculations.

Velocity factors \begin{equation}
f=\frac{c+v}{c-v}\label{eq:velocity-factor-def}\end{equation}
 and their square roots\begin{equation}
k=\sqrt{\frac{c+v}{c-v}}\label{eq:Bondi-k-def}\end{equation}
can be interpreted respectively as two-way and one-way doppler factors.
It is natural to seek a physical interpretation for rapidity as well.

Elegantly, L{\'e}vy-{L}eblond has interpreted rapidity as the integral
of proper acceleration\cite{Lej80}. Beginning at rest in a certain
frame, accelerating arbitrarily along a straight line, one's current
rapidity is always the integral of what an ideal accelerometer would
measure. If one chooses the arbitrary constant $b$ to be equal to
$c,$ the rapidity is the velocity that an ideal Newtonian inertial
guidance system would impute.

\section{The change-in-pitch interpretation of rapidity}

But we can also give another intepretation. From the definition of
rapidity, we find that \begin{eqnarray}
\alpha & = & b\arctanh\frac{v}{c}\nonumber \\
 & = & b\ln\sqrt{\frac{1+v/c}{1-v/c}}\nonumber \\
 & = & b\ln\sqrt{\frac{c+v}{c-v}}\nonumber \\
 & = & b\ln k,\label{eq:alpha-prop-to-ln-k}\end{eqnarray}
where $k=\sqrt{\frac{c+v}{c-v}}$ is the one-way doppler factor for
frequencies, for a velocity of approach $v$ between source and observer.
Rapidities compose by addition because they are logarithms of doppler
factors, and doppler factors compose by multiplication.

Using generic logarithms lg(.)\cite{Wia07angle-loglevel-generic-logs}
and taking $b=1\,\textrm{neper}=\lg\textrm{e},$ \begin{eqnarray}
\alpha & = & \ln k\,\lg\textrm{e}\nonumber \\
 & = & \frac{\lg k}{\lg\textrm{e}}\lg\textrm{e}\nonumber \\
 & = & \lg k.\label{eq:alpha-is-lg-k}\end{eqnarray}

This a very simple relationship, and says that the rapidity so defined
is just the logarithmic level of $k.$ We can further write\begin{eqnarray}
\alpha & = & \frac{\lg(k)}{\lg(2)}\lg(2)\nonumber \\
 & = & \log_{2}(k)\,\textrm{octaves}\nonumber \\
 & = & 12\log_{2}(k)\,\textrm{semitones},\label{eq:rapidity-as-change-in-pitch}\end{eqnarray}

for $k$ the doppler factor of frequency shift along the line of motion.
Human pitch perception is logarithmic in frequency. If we had the
ability to hear electromagnetic radiation, the rapidity $\alpha$
is simply the doppler-induced shift in pitch. The pitch of radiation
from the direction boosted toward goes sharp by $\alpha,$ while the
pitch from the opposite direction goes flat by $\alpha.$

\section{Discussion}

The following are all equivalent one-dimensional kinematic states
:

\begin{itemize}
\item light from directly ahead is sharp---blueshifted---by one semitone,
\item light from directly behind is flat---redshifted---by one semitone,
\item proper acceleration integrated with respect to proper time imputes
a Newtonian velocity of $c\ln2^{1/12}\approx0.057762\, c,$ but
\item the relativistically correct velocity is \begin{equation}
\frac{(2^{1/12})^{2}-1}{(2^{1/12})^{2}+1}c\approx0.057698\label{eq:rel-correct-vel}\end{equation}

\end{itemize}
At nearly 6\% of $c,$ the rapidity is one semitone, while the relativistically
correct velocity differs from the Newtonian calculated velocity by
about one part in a thousand. For many purposes, then, below this
speed one can treat the rapidity and velocity as proportional, and
even identify them. On the other hand, the semitone is still a large
rapidity compared to those usual for macroscopic terrestrial objects.
Taking the fine structure constant as 1/137 and the mass of electron
to be negligible compared to that of the proton, then the velocity
of an orbital electron in a hydrogenic atom is $c/137.$ This velocity
corresponds to a rapidity of about an eighth of a semitone. The cent,
a pitch unit used in piano tuning and defined as one percent of a
semitone, corresponds to a velocity of \begin{equation}
\frac{2^{2/1200}-1}{2^{2/1200}+1}c\approx\ln2\cdot c/1200\approx173\,\textrm{km/s},\label{eq:cent-as-a vel}\end{equation}

still a very large velocity for macroscopic terrestrial objects. For
comparison, the orbital velocity of the sun about the galactic center
is about 217 km/s, escape velocity at the surface of the sun is about
618 km/s, and escape velocity from Jupiter's surface is about 60 km/s.

A rapidity of one millionth of a semitone corresponds to a speed of
about 17.32 m/s, which is just over 60 km/h, or just under 40 miles/hour.
Highway patrol radar guns need to resolve frequency differences to
at least an order of magnitude smaller, or a ten millionth of a semitone.

One nanosemitone corresponds to about 1.732 cm/s. Still, these quantities
differ in much the way that a small angle differs from its tangent.
Rapidities add in a relavistically correct fashion, while velocities
do not, except approximately. If human vision had such extreme frequency
resolution that we could see doppler effects at familiar speeds, or
if we could somehow hear light to such resolution, perhaps an intuitive
understanding of relativity would be much easier to develop.

\bibliographystyle{plain}

\end{document}